\pdfoutput=1
\PassOptionsToPackage{pagebackref}{hyperref}
\documentclass[sigconf,nonacm,pbalance]{acmart-arxiv}
\usepackage{graphicx}
\usepackage{booktabs}
\usepackage{tabularx}
\usepackage{colortbl}
\definecolor{tableheader}{HTML}{D9D9D9}
\usepackage{algorithm}
\usepackage[noend]{algpseudocode}
\usepackage{microtype}
\usepackage{placeins}
\usepackage{flafter}
\hypersetup{colorlinks=true,linkcolor=blue,citecolor=blue,urlcolor=blue}
\makeatletter
\newcommand*\NTcite@first[1]{%
  \ifcsname NTcite@seen@#1\endcsname
  \else
    \global\expandafter\let\csname NTcite@seen@#1\endcsname\@empty
    \Hy@raisedlink{%
      \hyper@anchorstart{nt.firstcite.#1}\hyper@anchorend}%
  \fi}
\newcommand*\NTcite@label[2]{%
  \ifcsname NTcite@seen@#1\endcsname
    \begingroup
      \hypersetup{hidelinks}%
      \hyperlink{nt.firstcite.#1}{\@biblabel{#2}}%
    \endgroup
  \else
    \@biblabel{#2}%
  \fi}
\patchcmd{\NAT@citexnum}
  {\NAT@parse{\@citeb}}
  {\NAT@parse{\@citeb}\NTcite@first{\@citeb\@extra@b@citeb}}
  {}{\PackageError{citation-links}{Cannot patch natbib numeric citations}%
    {The natbib citation implementation has changed; review this patch.}}
\patchcmd{\NAT@citex}
  {\NAT@parse{\@citeb}}
  {\NAT@parse{\@citeb}\NTcite@first{\@citeb\@extra@b@citeb}}
  {}{\PackageError{citation-links}{Cannot patch natbib author-year citations}%
    {The natbib citation implementation has changed; review this patch.}}
\patchcmd{\NAT@anchor}
  {\@biblabel{#2}}
  {\NTcite@label{#1\@extra@b@citeb}{#2}}
  {}{\PackageError{citation-links}{Cannot patch natbib bibliography labels}%
    {The natbib bibliography implementation has changed; review this patch.}}
\makeatother

\renewcommand*{\backref}[1]{}
\renewcommand*{\backrefalt}[4]{%
  \ifcase #1\relax\or\space(Cited on p.~#2.)\else\space(Cited on pp.~#2.)\fi}
\setcopyright{none}
\acmDOI{}
\acmISBN{}
\acmYear{2027}
\copyrightyear{2027}
\acmConference[ICPE '27]{ACM/SPEC International Conference on Performance Engineering}{2027}{}
\title[High-Throughput Serving of Nemotron 3.5 Transcribe]{High-Throughput Serving of Nemotron 3.5 Transcribe with Regime-Aware Megakernels and Kernel Fusion}
\author{Aditya Chatterjee}
\affiliation{\institution{NVIDIA}\city{Bangalore}\country{India}}
\email{adichatterje@nvidia.com}
\author{Sanjay Singh Chauhan}
\affiliation{\institution{NVIDIA}\city{Pune}\country{India}}
\email{schauhan@nvidia.com}
\author{Mayank Jain}
\affiliation{\institution{NVIDIA}\city{Pune}\country{India}}
\email{mayjain@nvidia.com}
\author{Arushi Dhingra}
\affiliation{\institution{NVIDIA}\city{Pune}\country{India}}
\email{arushid@nvidia.com}
\author{Sahin Chowdhury}
\affiliation{\institution{NVIDIA}\city{Pune}\country{India}}
\email{sahinc@nvidia.com}
\author{Rahul Mittal}
\affiliation{\institution{NVIDIA}\city{Pune}\country{India}}
\email{rmittal@nvidia.com}
\renewcommand{\shortauthors}{Chatterjee et al.}

\makeatletter
\def\@mkauthors@iii{%
  \global\setbox\mktitle@bx=\vbox{%
    \unvbox\mktitle@bx\par\medskip
    \centering\normalfont\normalsize
    Aditya Chatterjee\textsuperscript{1}, Sanjay Singh Chauhan\textsuperscript{2},
    Mayank Jain\textsuperscript{2}, Arushi Dhingra\textsuperscript{2},
    Sahin Chowdhury\textsuperscript{2}, Rahul Mittal\textsuperscript{2}\par\smallskip
    \textsuperscript{1} NVIDIA, Bangalore, India\par
    \textsuperscript{2} NVIDIA, Pune, India\par\smallskip
    \{\href{mailto:adichatterje@nvidia.com}{adichatterje},
    \href{mailto:schauhan@nvidia.com}{schauhan},
    \href{mailto:mayjain@nvidia.com}{mayjain},
    \href{mailto:arushid@nvidia.com}{arushid},
    \href{mailto:sahinc@nvidia.com}{sahinc},
    \href{mailto:rmittal@nvidia.com}{rmittal}\} @nvidia.com\par\bigskip}}
\makeatother
\begin{document}
\begin{abstract}
Nemotron 3.5 Transcribe ranks first among open-weight models on the Hugging Face Open ASR leaderboards. Serving it efficiently takes more than fast operators: continuous batching changes the amount and mix of work at every iteration, and the best fusion granularity changes with it. We present a serving engine that, for every pass, chooses between persistent megakernels and chains of fused kernels at crossovers measured offline. On an H100 SXM GPU, throughput of our accuracy-preserving serving peaks at 4,345\(\times\) real time, 1.39\(\times\) the peak of tuned vLLM, and is up to 4.39\(\times\) faster at low concurrency; on B200 it peaks at 6,367\(\times\), 1.64\(\times\) vLLM at the highest concurrency we measured. Word error rate (WER) is 1.91 to 1.93\% against 1.88 to 1.93\% for vLLM. Within a 400 ms budget for p95 end-to-end completion time, it serves 8\(\times\) as many concurrent requests. Choosing per pass beats either level alone: 23\% higher throughput than always using megakernels at 128 concurrent requests, and 9\% higher than always using fused kernels at 8. Opt-in switches, including greedy MoE layer skipping, add 13 to 16\% throughput on B200 for offline jobs that can accept higher WER.

\end{abstract}
\maketitle
\fancyhead{}
\section{Introduction}\label{sec:intro}

Speech transcription is a core operation in speech-to-speech (S2S) systems and modern user-facing applications. Nemotron 3.5 Transcribe is the leading open speech transcription model at the time of its release in October 2026. It combines speech and speaker encoders with a hybrid Mamba Transformer mixture-of-experts (MoE) based language model, supports transcription in 40 languages and processes recordings up to one hour long.

Although inference serving frameworks like vLLM~\cite{vllm} provide optimized kernels, continuous batching and speculative decoding, vLLM's existing Nemotron 3.5 Transcribe integration leaves significant opportunities for optimization across operator boundaries.

Separate kernels for normalization, quantization, projections, expert routing and output processing introduce intermediate memory traffic and synchronization. Fusion can reduce these costs but constrain registers, occupancy and compute efficiency. The challenge is to choose fusion granularity as serving workload size changes.

\begin{figure}[!htb]
\centering
\includegraphics[width=2.50in]{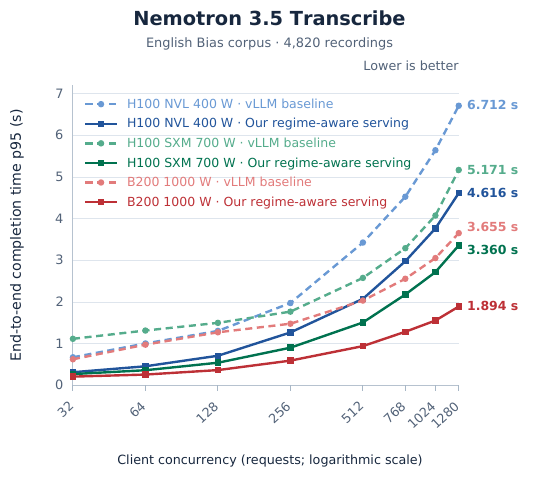}
\caption{p95 end-to-end completion time across client concurrency.}\label{fig:latency}
\Description{Completion time versus client concurrency for vLLM and the regime-aware engine on H100 NVL, H100 SXM, and B200. The optimized engine has lower p95 completion time across all plotted concurrency levels.}
\end{figure}

We address this problem with a regime-aware serving design which reduces p95 end-to-end completion time by 1.50 to 4.10\(\times\) on H100 SXM, 1.45 to 2.19\(\times\) on H100 NVL and 1.93 to 3.75\(\times\) on B200 across 32 to 1,280 concurrent requests (Figure~\ref{fig:latency}). Our approach combines persistent megakernels, which execute a model region within one launch, with specialized kernel chains that fuse selected operations while retaining independent launch configurations. We select between these implementations using measured behavior across token lengths. The resulting native serving engine also processes prompt and verification work together, adapts expert computation to routed populations and keeps speculative drafting and token selection on the GPU.

Our major contributions include:

\begin{itemize}
\item A design methodology for prefill-dominated, megakernel-receding architectures using regime-aware megakernels and kernel fusion.

\item An engine inside a production inference server that, for every pass, chooses between megakernels and fused kernels at measured crossovers.

\item Regime-aware kernels designed for Nemotron 3.5 Lightning apply to other models built on it.

\item Optimizations for applications that can tolerate a higher word error rate (WER), including greedy selection of MoE layers to skip within a WER budget.

\end{itemize}

\section{Industrial context}\label{sec:context}

As of 6 November 2026, it ranks first among open-weight models on the English, multilingual and long-form Hugging Face Open ASR leaderboards~\cite{open-asr,nemotron-transcribe-launch} (5.33\%, 4.62\% and 10.98\% WER) and first among 14 systems on VoiceArena's speaker-attributed transcription benchmark~\cite{voicearena}.

\subsection{Nemotron 3.5 Transcribe model}\label{sec:model}

Nemotron 3.5 Transcribe \cite{nemotron-transcribe} is NVIDIA's language model for transcription. One model, steered by a natural-language instruction, produces verbatim or written-form, contextual, speaker-attributed and rich transcripts, translates speech and identifies the spoken language in 40 languages.

\begin{figure}[!htb]
\centering
\includegraphics[width=2.55in]{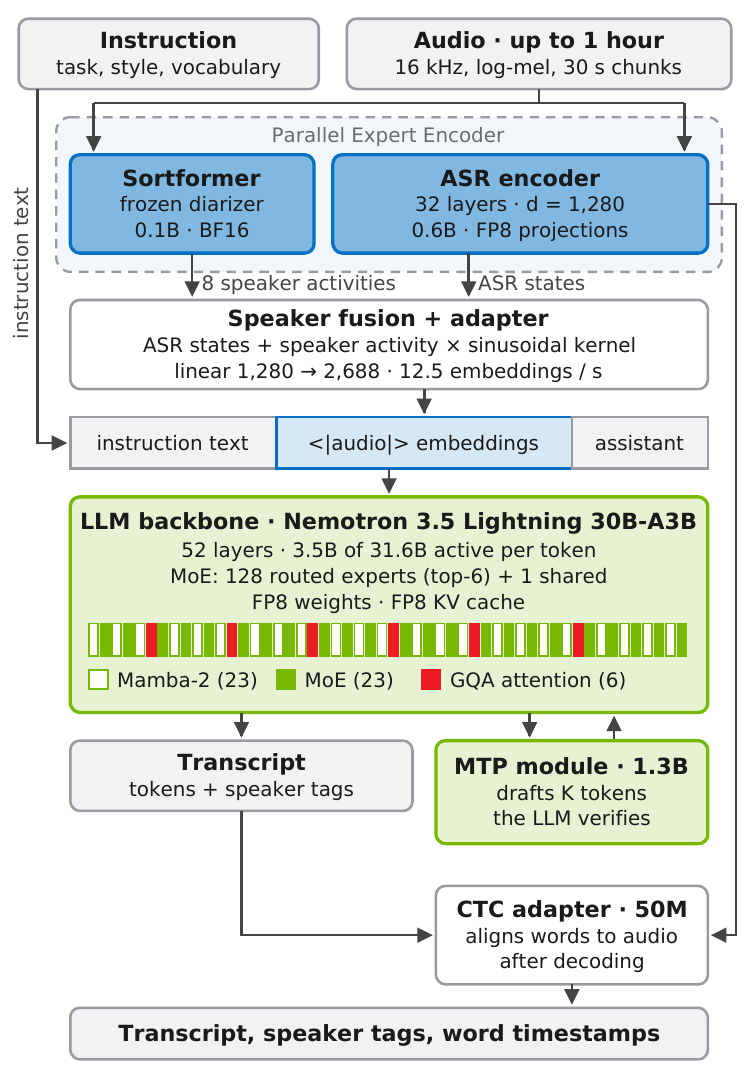}
\caption{Nemotron 3.5 Transcribe model architecture.}\label{fig:model}
\Description{Audio flows through parallel Sortformer and ASR encoders, speaker fusion, the modality adapter, and the hybrid Mamba, MoE, and GQA language model. MTP drafts tokens, and a CTC adapter aligns words to audio.}
\end{figure}

Its audio path is a Parallel Expert Encoder: a 32-layer Transformer ASR encoder (d=1,280, 0.6B parameters) and a frozen Sortformer~\cite{sortformer} diarization model (0.1B) run side by side on 128-bin log-mel features and the per-frame activities of up to eight speakers are added to the ASR states through a fixed sinusoidal kernel. A linear adapter projects the result to the LLM width (1,280 \(\rightarrow\) 2,688). The LLM backbone, Nemotron 3.5 Lightning 30B-A3B, has 52 layers: 23 Mamba-2~\cite{mamba2}, 23 MoE layers with 128 routed experts (top-6) and a shared expert and only 6 grouped-query attention (GQA) layers so 3.5B of its 31.6B parameters are active per token. A multi-token prediction (MTP) module (1.3B parameters) serves as the draft model for speculative decoding and a CTC~\cite{ctc2006} adapter (50M parameters) on the ASR encoder aligns words to audio. The complete model contains approximately 34B parameters, with approximately 4.3B active parameters including the audio-processing components (Figure~\ref{fig:model}).

\subsection{Target workload and serving requirements}\label{sec:workload}

We target transcription at high concurrency. Our benchmark is NVIDIA's English Bias corpus (test split): 4,820 single-speaker recordings totaling 15.36 hours (3.5 to 58.8 s, 11.5 s on average, 25 words per transcript on average). The recordings are split evenly across four domains (telecom and finance, technical, alphanumeric codes and numbers, and general) and six US and UK accents, with an even gender split. Our primary evaluation uses the full Bias corpus; we additionally evaluate the complete LibriSpeech~\cite{librispeech} test-clean and test-other splits. Expert and layer rankings for the opt-in mode (Section~\ref{sec:opt-in}) come from a separate calibration set of 512 LibriSpeech dev utterances.

We report end-to-end RTFx, the seconds of audio transcribed per second over a whole pass including ramp-up and drain; latency per request (p95); and WER and character error rate (CER). We measure on single GPUs: H100 SXM (700 W), H100 NVL (400 W) and B200 (144 SMs, 1,000 W). Every comparison is a matched run on the same GPU.

Online serving maximizes concurrent requests while keeping p95 end-to-end completion time below 400 ms, within a 1-second S2S budget allocating 100 ms to TTS and 500 ms to the LLM. Offline serving maximizes RTFx across batch sizes.

\subsection{Deployment and use cases}\label{sec:deployment}

Nemotron 3.5 Transcribe was released as open weights on Hugging Face on 28 October 2026 under the OpenMDW-1.1 license, with BF16, FP8 and NVFP4 checkpoints. vLLM, SGLang, Hugging Face Transformers and NeMo-Speech.cpp support it, and inference providers including Baseten, DeepInfra and Simplismart have hosted it since release day. NVIDIA's reference service is a Rust inference server with OpenAI-compatible routes in front of a vLLM engine. It is packaged per GPU family, from Jetson Thor and DGX Spark to H100/H200 and GB200/GB300. Our engine runs in process inside that server through a C ABI with no Python in the serving path. It keeps the same API, word timestamps and speaker labels, and has parity tests against the vLLM reference. Our optimized serving engine will be available in a future release of NVIDIA Nemotron 3.5 Transcribe NIM~\cite{nemotron-transcribe-nim}.

One checkpoint serves two workloads. Offline, it transcribes call-center, meeting and media archives of up to an hour per recording, where transcription throughput (RTFx) decides deployment. Online, it is the recognizer in voice agents and speech-to-speech systems, where a short utterance must be transcribed within about 400 ms.

\section{Regime-aware megakernel architecture}\label{sec:architecture}

The vLLM baseline executes Nemotron 3.5 Transcribe at operator granularity: each operator, or a small group fused by torch.compile, is a separate kernel, launched by vLLM's model runner or replayed from a CUDA graph, and consecutive kernels pass intermediate results through global memory. Our optimized engine replaces every component with one of two fusion levels:

\begin{itemize}
\item A megakernel runs the whole region in one persistent launch.

\item Fused kernels run the region as a short chain of custom kernels, each fusing several operators, as CUTLASS~\cite{cutlass} GEMM epilogues and chained with programmatic dependent launch (PDL)~\cite{cuda-guide}.

\end{itemize}

\subsection{Why is one megakernel not enough?}\label{sec:why}

Prior work on megakernel composition \cite{chatterjee2026megakernel} found that a single megakernel is rarely the fastest design. That work describes each fused region by how its gain from a megakernel changes with batch size. A megakernel-stable region gains at every size. A megakernel-receding region gains up to a crossover and loses beyond it.

With in-flight batching, the batch changes at every iteration; we size each region by the work of one pass: tokens per encoder pass or rows per decoder step.

For a region which recedes beyond a crossover point, the engine keeps both a megakernel and a conventional implementation and switches between them. Conventional means one kernel launch per operation inside our engine, with CUTLASS or cuBLASLt GEMMs and our custom CUDA kernels.

Table~\ref{tab:regimes} lists the result. Two megakernels are stable and always run. Five are receding: three switch at runtime and two are selected by the serving configuration.

\begin{table}[!htb]
\caption{Megakernel regimes and selection thresholds.}\label{tab:regimes}
\centering
\small
\setlength{\tabcolsep}{4pt}
\setlength{\arrayrulewidth}{0.5pt}
\renewcommand{\arraystretch}{1.08}
\setlength{\extrarowheight}{1pt}
\begin{tabularx}{\linewidth}{|>{\raggedright\arraybackslash}p{.21\linewidth}|>{\raggedright\arraybackslash}p{.16\linewidth}|>{\raggedright\arraybackslash}X|}
\hline
\rowcolor{tableheader}
\textbf{Megakernel} & \textbf{Class} & \textbf{Selection threshold} \\ \hline
ASR encoder & Receding & T = 384 tokens per encoder pass (input T \(\times\) 1,280) (runtime) \\ \hline
Sortformer & Receding & T = 384 tokens per encoder pass (input T \(\times\) 512) (runtime) \\ \hline
Prefill & Receding & Prompt-only execution; at most 64 requests and 2,048 packed tokens per launch. Configuration-dependent availability. \\ \hline
Decoder - Mamba & Receding & Verification specialization supporting at most four rows per request. \\ \hline
Decoder - GQA & Stable & Retained across the evaluated eligible shapes. \\ \hline
Decoder - MoE & Receding & R = 448 rows per decoder step (input R \(\times\) 2,688) (runtime) \\ \hline
MTP draft & Stable & Retained across the evaluated eligible shapes. \\ \hline
\end{tabularx}
\end{table}

\begin{table}[!htb]
\caption{Megakernel speedup over fused kernels on H100; values above 1\(\times\) favor megakernels.}\label{tab:crossovers}
\centering
\small
\setlength{\tabcolsep}{4pt}
\setlength{\arrayrulewidth}{0.5pt}
\renewcommand{\arraystretch}{1.08}
\setlength{\extrarowheight}{1pt}
\begin{tabularx}{\linewidth}{|>{\raggedright\arraybackslash}p{.27\linewidth}|>{\raggedright\arraybackslash}X|}
\hline
\rowcolor{tableheader}
Region & Pass size: speedup \\ \hline
ASR encoder (tokens) & 64: 1.8\(\times\) \(\cdot\) 128: 1.6\(\times\) \(\cdot\) 256: 1.25\(\times\) \(\cdot\) 384: 1.0\(\times\) \(\cdot\) 1,024: 0.79\(\times\) \\ \hline
MoE layer (rows) & 64: 1.16\(\times\) \(\cdot\) 128: 1.12\(\times\) \(\cdot\) 256: 1.07\(\times\) \(\cdot\) 448: 1.0\(\times\) \(\cdot\) 1,024: 0.88\(\times\) \\ \hline
\end{tabularx}
\end{table}

\subsection{Nemotron 3.5 Transcribe is a prefill-dominated megakernel-receding architecture}\label{sec:receding}

The five megakernels, classified as megakernel-receding, dominate the end-to-end computation time, which makes the Nemotron 3.5 Transcribe model a megakernel-receding architecture.

The thresholds were measured offline and are applied from the pass size alone. At B128 most passes are large, so the encoder and MoE take the conventional path, while the GQA and draft megakernels run on every pass. At low load, and while the batch fills or drains, the megakernels take over.

Neither level alone matches this choice (Table~\ref{tab:crossovers}). Megakernels slow down large passes (7.66 against 6.05 ms per encoder pass at 1,024 tokens), and fused kernels slow down small ones (549 launches instead of four per encoder pass). A megakernel-only design would also need separate prompt steps, because the prefill megakernel accepts only prompt-only batches, so each iteration would stream the 29.8 GB of FP8 expert weights twice.

The workload is prefill-dominated at serving loads. Prefill rows consume 24.1\% to 46.2\% of end-to-end time across batch sizes 1 to 512. Decode rows dominate prefill rows only at batch size 1 where decode consumes 35.3\% while prefill is at 24.1\%.

\begin{figure*}[!htb]
\centering
\includegraphics[width=\textwidth]{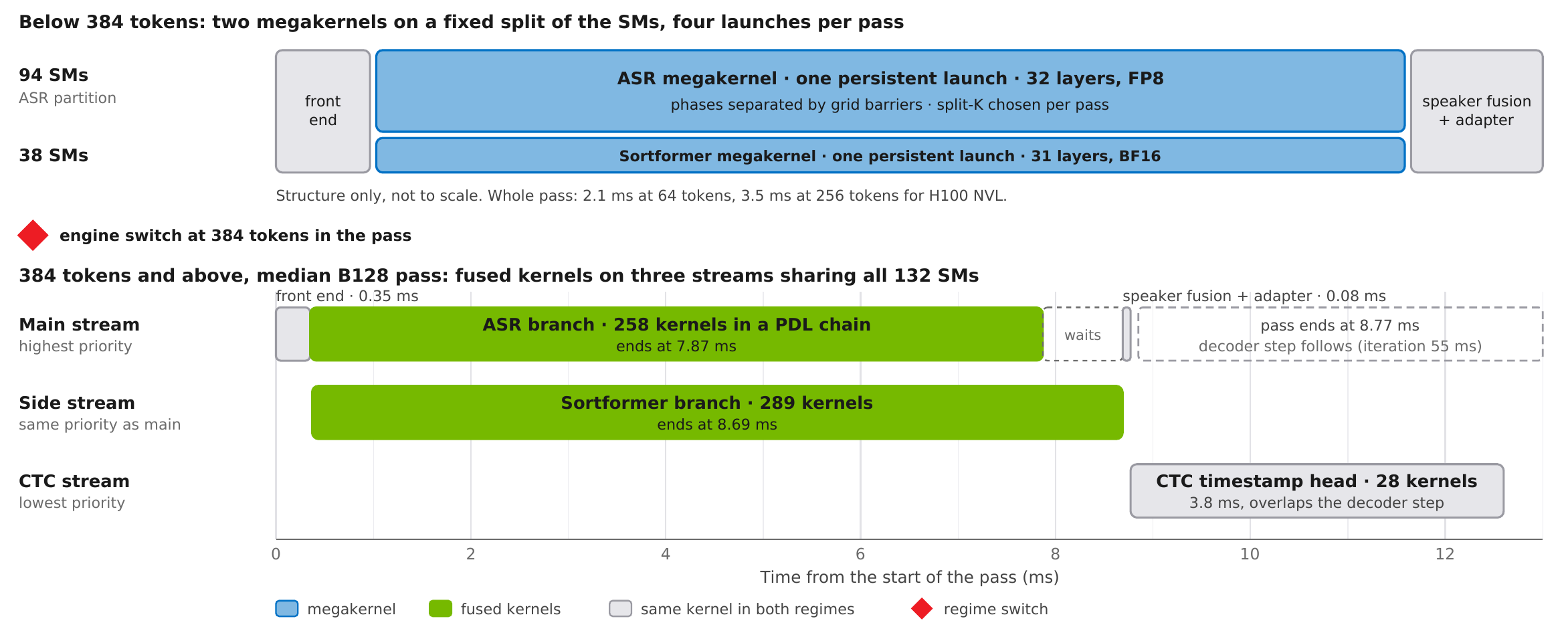}
\caption{Encoder execution across the 384-token crossover; CTC span measured during B128 serving (median 1,690 tokens per pass).}\label{fig:encoder-pass}
\Description{Small encoder passes use concurrent ASR and Sortformer megakernels with a fixed SM split. Large passes use fused kernels on concurrent streams, with the CTC timestamp head overlapping the decoder.}
\end{figure*}

\section{Speech encoder across load regimes}\label{sec:encoder}

In the speech encoder, each iteration encodes the audio of all requests it admits in one pass. One second of audio becomes 12.5 encoder tokens, so the size of a pass is the amount of audio admitted in that iteration.

\subsection{One encoder pass}\label{sec:encoder-pass}

Figure~\ref{fig:encoder-pass} shows a pass in B128 serving. A front-end kernel computes log-mel features for all requests. The two branches of the Parallel Expert Encoder \cite{nemotron-transcribe} run concurrently: the ASR branch (32 layers, d = 1,280, FP8 projections) on the main stream and the Sortformer diarization branch (31 layers, d=512, BF16) on a side stream. A tail kernel adds the speaker activities to ASR states and projects the result to the LLM width (modality adapter). The CTC timestamp head runs last, on the lowest-priority stream, overlapping the decoder step; its output feeds the draft hints.

The pass is queued behind the previous iteration's MTP draft, so the host builds the next batch while the GPU encodes. In steady-state B128 serving, the 2,048-row prompt budget of a decoder step limits a pass to about 1,800 tokens: passes carry 1,218 to 1,798 tokens (median 1,690 or 135 s of audio) and take 8.8 ms of the 55 ms iteration on H100 NVL. Under light load, a pass holds one or a few requests; a 5- to 10-second utterance is 63 to 125 tokens.

\subsection{Megakernels for small passes}\label{sec:small-passes}

We have designed two separate megakernels for ASR and Sortformer. Below 384 tokens, the ASR and Sortformer megakernels launch on separate streams with persistent grids of 94 and 38 CTAs, respectively.

In every CTA, a producer warp loads tiles of weights and activations with TMA into a six-stage shared-memory ring and two consumer warpgroups multiply them with wgmma instructions~\cite{ptx-isa}, in FP8 for ASR and BF16 for Sortformer.

\begin{figure}[!htb]
\centering
\includegraphics[width=2.90in]{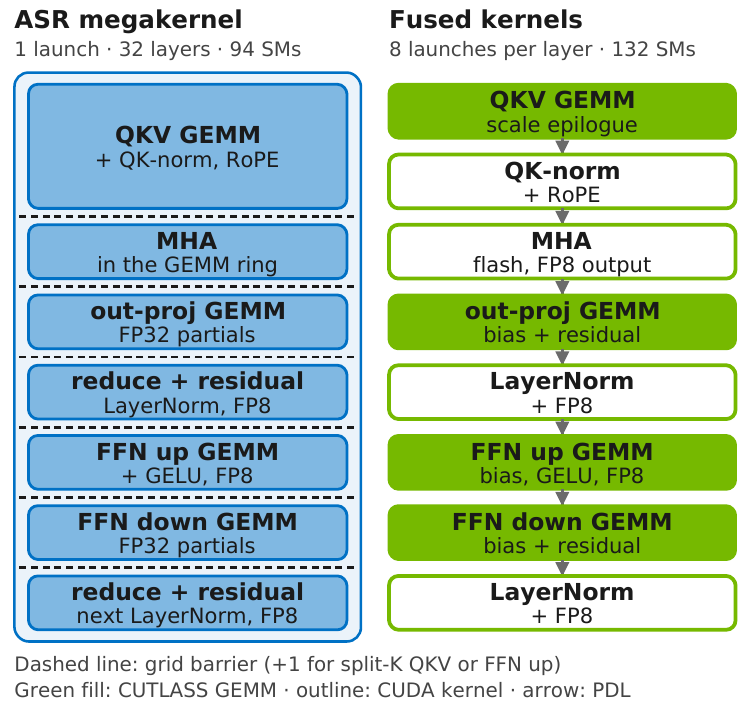}
\caption{ASR layer phases in the megakernel and fused-kernel paths.}\label{fig:asr-phases}
\Description{The megakernel separates phases with grid barriers. The fused path implements QKV projection, normalization and RoPE, attention, output projection, and feed-forward phases as eight kernels.}
\end{figure}

Each layer is a sequence of phases separated by software grid barriers. Figure~\ref{fig:asr-phases} illustrates the phases for ASR. The Sortformer megakernel follows the same plan and ends with the speaker head.

Small passes have too few output tiles to fill the GPU. At 64 tokens, the projections expose 10 to 40 output tiles. Under a one-working-CTA-per-SM mapping, this provides useful tile work for approximately 8 to 30\% of 132 SMs. Split-K tile scheduling increases the available parallel work to 50 to 80 tasks, corresponding to approximately 53 to 85\% of the 94 CTA ASR worker pool.

Each phase boundary costs a grid barrier instead of a kernel launch, and a whole pass takes four launches instead of 549. We synchronize at phase boundaries instead of scheduling tiles across operators. In tile-level dataflow, as in recent LLM megakernels \cite{mpk2026,no-bubbles}, each tile of the next operator waits on counters for only the tiles it reads, so consecutive phases can overlap. Our encoder passes leave little to overlap. Below 384 tokens a pass has at most three 128-row blocks, so each phase fits in one or two waves of tiles on 94 SMs, and every phase reads whole rows: LayerNorm needs all ten column tiles of a row and their split-K partial sums, GQA needs every key and value of a request, and the down-projection needs all 40 tiles of the 5,120-wide FFN row. The tiles that one next-phase tile reads are spread over most CTAs and finish at about the same time, so per-tile counters would start the next phase little earlier than a barrier, while adding release and acquire cost to every tile. Based on experiments, a megakernel with tile-level scheduling is 4 to 10\% slower at every B128 shape than the megakernel with grid barriers: its dependency fences added latency to every task.

\subsection{Kernel fusion for large passes}\label{sec:large-passes}

Beyond 384 tokens, each ASR layer runs as eight kernels (Figure~\ref{fig:asr-phases}). The four projections for the H100 GPU are CUTLASS SM90 FP8 GEMMs (TMA, warp-specialized, 128\(\times\)128\(\times\)128 tiles) whose epilogue visitor trees apply the per-channel weight scales: the output and FFN down-projections read the residual stream and add bias and residual in place, and the FFN up-projection applies bias, GELU and FP8 quantization for the next GEMM. Four custom kernels complete the layer: QK LayerNorm with RoPE, a variable-length FlashAttention-3~\cite{flashattention3} that writes FP8 for the output projection and two LayerNorm-plus-FP8 kernels. The ASR kernels launch with programmatic dependent launch (PDL): each kernel is launched while its producer still runs and waits for the producer's results on the GPU so the chain of 258 kernels per pass has significantly fewer launch gaps.

The Sortformer branch runs as 289 kernels: cuBLASLt BF16 GEMMs, row kernels for RoPE, attention, bias with residual and LayerNorm, and GELU, and the speaker head. Instead of a fixed SM split, the two branches share all SMs: they are queued layer by layer on two streams of equal priority, so each branch fills the SMs the other leaves idle.

The front end and the speaker fusion are one fused kernel each, in both regimes.

\section{Nemotron 3.5 Lightning backbone}\label{sec:backbone}

Prefill and decode phases share the same Nemotron 3.5 Lightning backbone which results in 45\% to 61\% of the end-to-end computation time being consumed by this backbone.

\subsection{Prefill and decode share one backbone}\label{sec:shared-backbone}

The backbone does two jobs:

\begin{itemize}
\item Prefill processes a new request's prompt once (the instruction, 85 audio embeddings and a short text suffix) to fill the request's Mamba states and KV cache and produce its first token: about 99 rows for a typical 6.8 s request after the shared prefix is reused.

\item Decode extends the transcript. With speculative decoding, each decode pass verifies the last accepted token plus up to 15 drafted tokens per request, about 9 rows on average, and a request needs three to four passes: about 30 decode rows for the same request.

\end{itemize}

The engine runs both jobs in the same step: each step reads every weight once, for the prefill rows of the requests admitted in that iteration and the decode rows of all requests in flight.

Each second of input audio gives the backbone about 21 prompt tokens to prefill (12.5 audio embeddings and, on average, 8.6 instruction and template tokens) but yields only about 3.6 transcript tokens, which speculative decoding verifies in about 4.4 decode rows. Even after prefix reuse, prompt rows are 76\% of the backbone's rows, and although a transcript token costs about four times as much backbone time as a prompt token (60 against 16 \(\mu\)s at B128), prefill takes 1.6 times as long as decode: 36\% against 22\% of serving time in tuned vLLM.

\subsection{Prefill megakernel design}\label{sec:prefill}

\begin{figure}[!htb]
\centering
\includegraphics[width=2.70in]{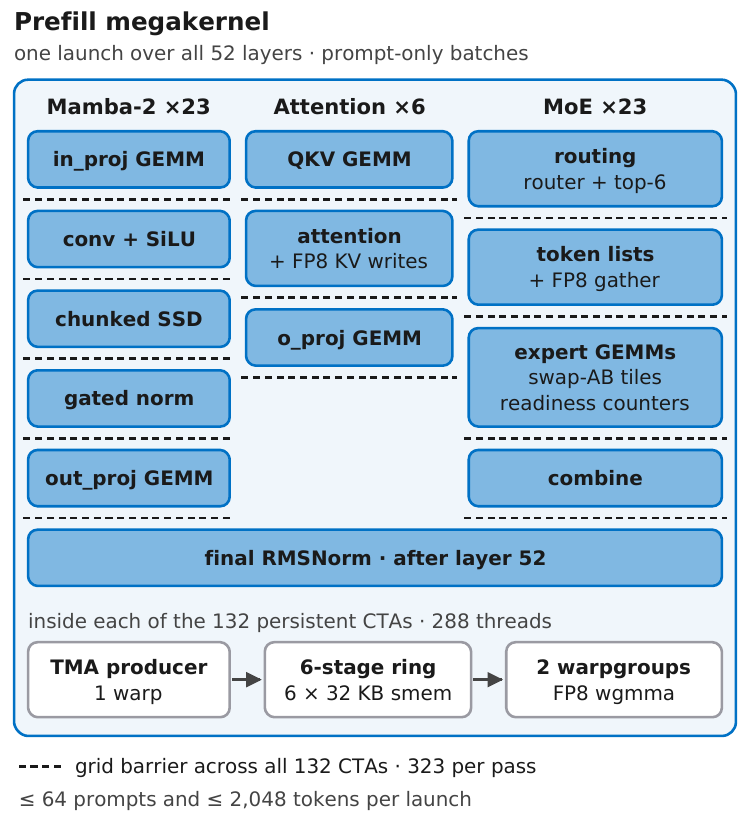}
\caption{Prefill megakernel phases and shared execution pipeline.}\label{fig:prefill}
\Description{A single cooperative launch processes Mamba, attention, and MoE layers using TMA producers, a shared-memory ring, and consumer warpgroups. Grid barriers separate the phases.}
\end{figure}

The prefill megakernel is one cooperative launch of 132 persistent CTAs, one per SM in the H100 GPU, that runs a packed batch of up to 64 prompts and 2,048 tokens through all 52 layers. The producer streams 128\(\times\)128 FP8 tiles of weights and activations from HBM into a six-stage shared-memory ring (6 \(\times\) 32 KB) with cp.async.bulk.tensor, and per-stage mbarriers carry the transaction byte counts. Its distinguishing choice is one pipeline for three operator families (Mamba-2, GQA and MoE) inside one launch (Figure~\ref{fig:prefill}).

Phases end at software grid barriers. For MoE, expert tiles are swap-AB: the weight takes wgmma's M side and the expert's 8 to 128 tokens its N side, chosen per expert, so small expert populations no longer pad to 128 rows. Per-expert readiness counters replace the barrier between up and down projections.

For a fixed cohort, all new incoming requests are computed using the prefill megakernel. In in-flight batching, prefill rarely runs alone as a monolithic prefill megakernel: at B128, 90\% of engine steps mix the prompt rows of newly admitted requests with the decode rows of running ones, 9\% carry only decode rows (the drain at the end of a pass) and 1\% carry only prompts. New prompts join the mixed step following the encoder pass and take the same kernels as the decode rows; that is the decoder's regime-aware design.

\subsection{Decoder regime-aware design}\label{sec:decoder}

\begin{figure}[!htb]
\centering
\includegraphics[width=2.55in]{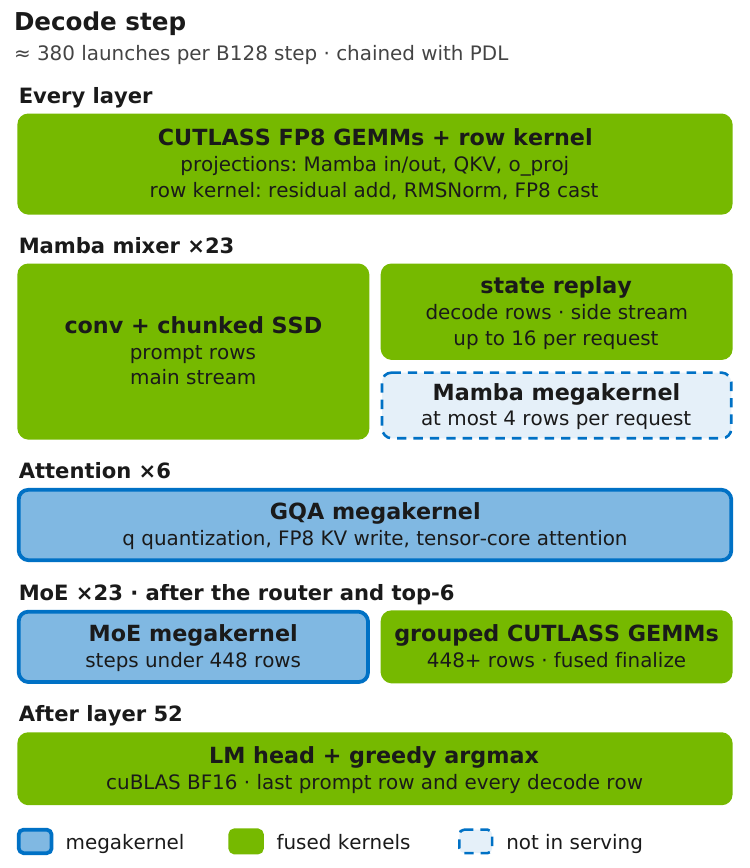}
\caption{Regime-aware kernel composition for the decoder step.}\label{fig:decoder}
\Description{Each decoder layer uses fused projections and row kernels. Mamba state replay and chunked SSD operate on separate streams; GQA uses a megakernel, and MoE selects a megakernel or grouped GEMMs according to pass size.}
\end{figure}

The decoder kernel is a combination of three megakernels and corresponding custom fused CUTLASS/cuBLAS kernels (Figure~\ref{fig:decoder}). During runtime, the GQA megakernel is executed for each iteration while the Mamba and MoE megakernels are replaced by corresponding custom fused kernels depending on the tensor dimensions.

The decode step launches its kernels layer by layer on a non-blocking stream at the highest priority. Our kernels and the CUTLASS GEMMs use programmatic dependent launch (PDL). The CUTLASS GEMMs are compiled with grid-dependency control. Prompt and decode rows share every weight read, so a B128 step streams its 29.8 GB of FP8 expert weights once for both kinds of rows. The dense projections are CUTLASS SM90 FP8 GEMMs (TMA, warp-specialized ping-pong mainloops, FP8 GEMMs with FP32 accumulator types, scale epilogues) and each layer starts with one row kernel for the residual add, RMSNorm and FP8 cast, which the MoE finalize absorbs after MoE layers. In a Mamba layer, the prompt rows run a fused convolution and chunk-parallel SSD on the main stream while a side stream, forked and joined with events, replays the decode rows: each request's 1 MB of BF16 state per layer is read into registers, up to 16 rows are verified from 16-entry x, dt and B rings on CUDA cores, and the state is written back only when the ring fills. The Mamba megakernel fuses the same replay into one launch but handles at most four rows per request.

The GQA megakernel quantizes queries, writes keys and values into the paged FP8 cache and computes attention with mma.sync m16n8k16 BF16 tiles whose 16-row M dimension holds the 16 query heads that share a KV head: each warp handles one query row for all 16 heads, and one CTA per GQA block and KV head stages the keys and values in shared memory once. In an MoE layer, a cuBLAS BF16 router and a top-6 kernel feed the MoE megakernel below 448 rows: one persistent launch that writes each token's FP8 copy straight to its six expert positions and the shared expert, runs both as swap-AB wgmma tiles of 8 to 64 tokens from nine 24 KB ring slots, and issues weight loads before a tile's inputs are ready. From 448 rows, CUTLASS ptr-array grouped FP8 GEMMs take over with the shared expert as a 129th group and ReLU\textsuperscript{2} and FP8 fused into the FC1 epilogue; above 1,200 rows they use 128 \(\times\) 128 tiles whose 2\(\times\)1 TMA multicast halves the token-tile reads, and they stream expert weights at 2.5 to 2.8 TB/s.

One finalize kernel combines the experts and applies the next layer's residual add, RMSNorm and FP8 cast, and the LM head is a cuBLAS BF16 GEMM with a greedy argmax kernel.

\section{Optimizing the draft model}\label{sec:draft}

The draft model for MTP is megakernel-stable.

\subsection{Draft model}\label{sec:draft-model}

The model's multi-token prediction (MTP) module has one attention block and one MoE block with shared weights, predicts the next token from the backbone's last hidden state and can be applied repeatedly. For each iteration, it proposes K tokens per request and the next decoder step verifies them.

Mean acceptance length \(\mathrm{MAL}(K)=1+\sum_{i=1}^{K}\prod_{j=1}^{i}p_j\) counts emitted tokens per verification for \(K\) drafts, including one bonus or correction token. Here, \(p_j\) is acceptance at position \(j\) conditional on preceding drafts' acceptance. Measured probabilities of 0.971, 0.977, 0.939, 0.931 and 0.931 yield \(\mathrm{MAL}(5)=5.41\) tokens.

\begin{figure}[!htb]
\centering
\includegraphics[width=2.60in]{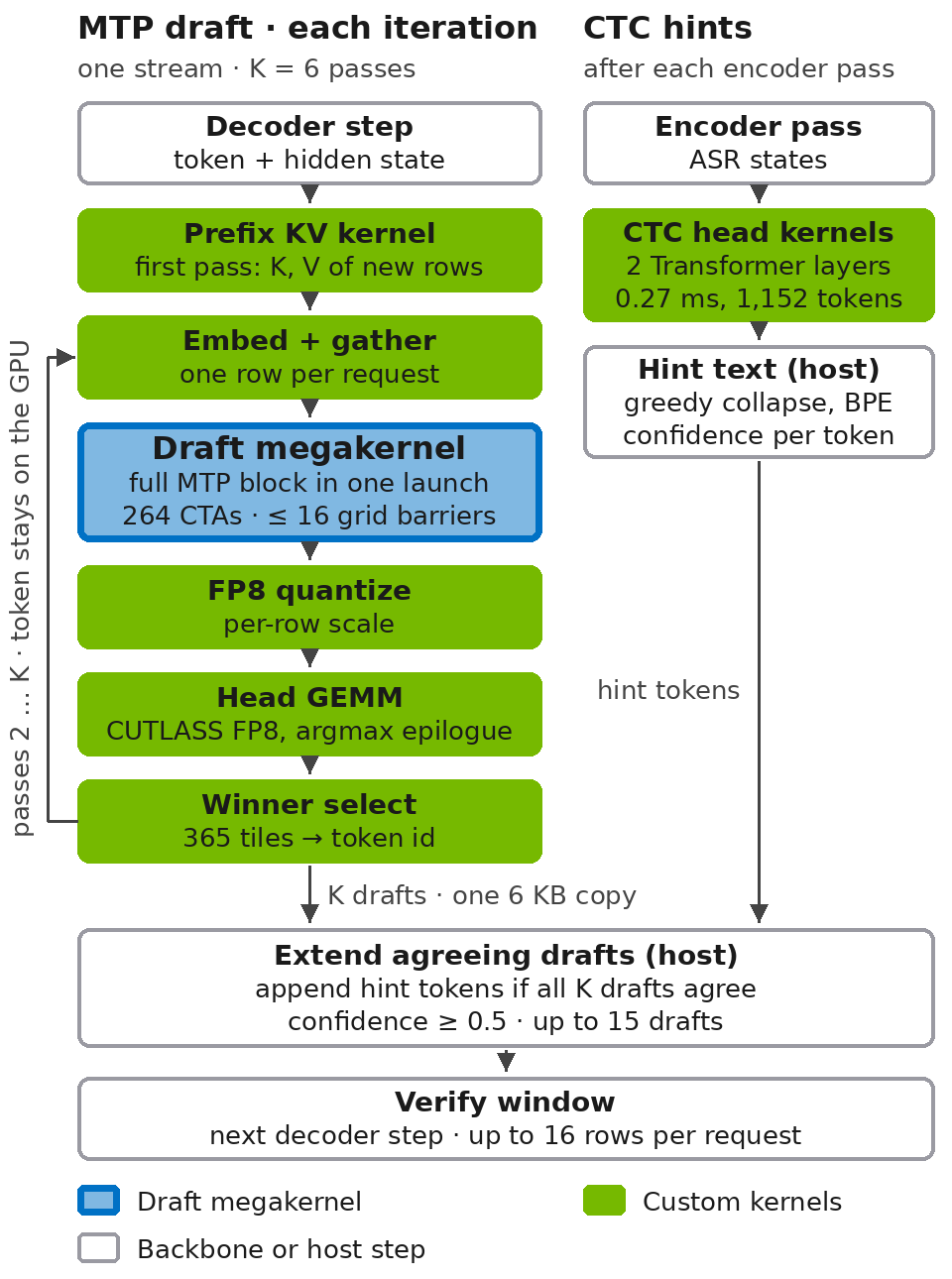}
\caption{MTP draft kernels and CTC hint flow; CTC timing uses a 1,152-token pass.}\label{fig:mtp}
\Description{The draft step combines prefix KV, embedding, draft megakernel, FP8 quantization, head GEMM, and winner-selection kernels. Encoder-side CTC hints extend agreeing drafts before verification.}
\end{figure}

We use one megakernel for the draft model which works in composition with other custom fused kernels (Figure~\ref{fig:mtp}).

\subsection{Draft megakernel design}\label{sec:draft-design}

The draft model runs as one megakernel. Each pass is one cooperative launch of 264 persistent CTAs, two per SM. Cooperative-groups grid barriers, up to 16 per pass, separate the phases of the MTP block.

The routed experts swap A and B: an expert's weights fill the 64-row M side and up to 32 of its tokens the N side, on an eight-stage ring with 64 \(\times\) 32 \(\times\) 16 wgmma; above 4 rows, the shared expert joins this GEMM as one more expert, which removes two phases and their barriers. Attention puts the 16 query heads of a KV head on the M=16 side of mma.sync m16n8k16: FP8 queries are converted straight into A-fragment registers, 32-key FP8 tiles of the cache are double-buffered into shared memory with cp.async and the CTA's four warps split the context and merge their softmax maxima and sums in shared memory; long contexts are split into KV segments that one more phase merges. Activations stay under 3 MB per buffer at 128 rows, so a pass's HBM traffic is its weights plus 512 B of cache per context token and request.

All drafts of an iteration, K\(\times\)128 token IDs, reach the host in one 6 KB copy at the end. The gaps between passes total 0.13 ms per iteration, so we kept one launch per pass.

\subsection{Custom fused kernels}\label{sec:draft-fused}

Six custom kernels feed and finish the megakernel, and the CTC head drafts from the encoder side (Table~\ref{tab:draft-kernels}). Three carry the main ideas: the prefix KV kernel computes only what the prompt rows need, the head GEMM reduces its logits to an argmax inside its epilogue, and the embedding kernel reads the head's token IDs in place, which keeps all K passes on the GPU. The head also never dequantizes: each input row is scaled so its absolute maximum maps to 448 (FP8 E4M3 maximum) and the scale is discarded, because a common positive scale across the vocabulary logits does not change their argmax.

\begin{table}[!htb]
\caption{Custom kernels surrounding the MTP draft megakernel.}\label{tab:draft-kernels}
\centering
\small
\setlength{\tabcolsep}{4pt}
\setlength{\arrayrulewidth}{0.5pt}
\renewcommand{\arraystretch}{1.08}
\setlength{\extrarowheight}{1pt}
\begin{tabularx}{\linewidth}{|>{\raggedright\arraybackslash}p{.27\linewidth}|>{\raggedright\arraybackslash}X|}
\hline
\rowcolor{tableheader}
\textbf{Kernel} & \textbf{Launch details} \\ \hline
Prefix KV copy & Grid: chunks \(\times\) tensors \(\times\) slots, 256 threads, PDL. \\ \hline
Embed + gather & rows \(\times\) 128 or rows \(\times\) 256 threads, PDL. One CTA per row. \\ \hline
Prefix KV kernel & 132 \(\times\) 288 threads, one CTA per SM, 203 KB shared memory, cooperative. Two consumer warpgroups and one TMA loader warp; five-stage ring of 40 KB stages. \\ \hline
FP8 input quantize & rows \(\times\) 256 threads, 32 B shared memory. \\ \hline
Head GEMM, argmax epilogue & 132 \(\times\) 384 threads, persistent, one CTA per SM. CUTLASS SM90 FP8 GEMM with a TMA warp-specialized mainloop and fast accumulation. \\ \hline
Winner select & \(\lceil\)rows / 32\(\rceil\) \(\times\) 256 threads, 2 KB shared memory. \\ \hline
CTC hint head & Row kernels \(\lceil\)tokens / 8\(\rceil\) \(\times\) 256; attention requests \(\times\) 16 \(\times\) \(\lceil\)frames / 64\(\rceil\) \(\times\) 128; lowest-priority stream. \\ \hline
\end{tabularx}
\end{table}

\section{What optimizations did not land?}\label{sec:optimizations}

\subsection{Optimizations impacting accuracy}\label{sec:opt-in}

As some optimizations trade WER for throughput, the engine exposes them only as opt-in features.

Expert masking removes the 32 least-routed experts per layer from routing, so their weights are never streamed (+3.0\% throughput at +4.2\% relative WER). Fewer experts per token means fewer routing rounds per row (top-5: +1.5\% at +1.7\%; top-4: +3.5\% at +6.2\%). Diarization bypass skips the Sortformer branch for single-speaker audio (+2.1\% with no WER change on LibriSpeech dev). Layer skipping removes whole MoE layers, chosen by greedy forward selection on 512 LibriSpeech dev utterances: each round adds the layer with the smallest WER increase and stops when calibration WER would exceed 2.55\%. It selects 7 of 23 MoE layers (6, 8, 13, 20, 27, 38 and 43) for 11\% more throughput; the Bias corpus is never used for selection.

We use the methodology in Algorithm~\ref{alg:skip-layers} to select which MoE layers to skip.
\begin{algorithm}[!htb]
\caption{\textsc{SkipLayers}$(L,D,X)$}\label{alg:skip-layers}
\small
\textbf{Inputs:} $L$ = $N$ MoE layers; $D$ = calibration set;\\
\hspace*{1em}$X$ = allowed WER increase over the unskipped model\\
\textbf{Output:} $S$ = layers to skip
\begin{algorithmic}[1]
\State $S \gets \emptyset$; $C \gets L$
\While{$C \ne \emptyset$}
  \State $H \gets$ empty min-heap ordered by error
  \For{each layer $l$ in $C$}
    \State $e_l \gets \mathrm{WER}_D(S \cup \{l\})-\mathrm{WER}_D(\emptyset)$
    \State push $(e_l,l)$ onto $H$
  \EndFor
  \State $(e^*,l^*) \gets \operatorname{pop}(H)$
  \State \textbf{if} $e^*>X$: \textbf{break}
  \State $S \gets S\cup\{l^*\}$
  \State $C \gets \emptyset$
  \While{$H\ne\emptyset$ and $\operatorname{top}(H).e\le X$}
    \State $(e,l) \gets \operatorname{pop}(H)$
    \State $C \gets C\cup\{l\}$
  \EndWhile
\EndWhile
\State \textbf{return} $S$
\end{algorithmic}
\end{algorithm}

Together the switches add 13 to 16\% throughput on B200 but raise WER from 1.92\% to 3.44 to 3.48\%, so they suit only cost-bound offline jobs; a light preset of diarization bypass and top-5 routing adds 3.6\% for 1.7\% relative WER.

\subsection{Optimizations that lowered performance}\label{sec:failed}

An expert-major MoE kernel, a persistent warp-specialized FP8 GEMM that kept token tiles resident, was 1 to 3\% slower than the grouped CUTLASS GEMMs, because the SM clock fell to 0.82 to 1.18 GHz while weights streamed under the 400 W cap in H100 NVL.

INT4 experts halved the expert bytes, but their mixed-input grouped GEMMs took 400 to 630 \(\mu\)s per FC1 call against 230 to 245 \(\mu\)s in FP8.

Green-context SM partitions between the draft and the encoder never beat sharing all 132 SMs (the best split, 1,818 RTFx, was within noise of 1,809).

Decode megakernels inside vLLM sped decode-only steps by 7.6 to 16.4\% but moved throughput by \(\times\)0.9965 because those steps carried 0.3\% of the tokens.

A faster Sortformer could not shorten encoder passes in which the ASR branch was longer, and larger encoder GEMM tiles left SMs idle with only 90 tiles at 1,152 tokens. Two experts per token reached 5.4\% WER and ran slower than three. Each failure traces to one of three principles, the lessons we would carry to another hybrid model:

\begin{enumerate}
\item Power wall: the board always draws 400 W, so time follows energy, \(t\approx(e_bB+e_f\Phi)/P\); fewer bytes or FLOPs shorten work and better scheduling mostly lowers the clock (expert-major kernel, INT4 experts, green contexts).

The classic roofline \cite{roofline} says a kernel needs at least the longer of its memory time and its compute time:

\[t\ge\max\!\left(\frac{B}{\mathrm{BW}_{\mathrm{peak}}},\frac{\Phi}{F_{\mathrm{peak}}}\right).\]
This assumes the GPU can run memory and arithmetic at their peaks simultaneously, so the shorter one hides under the longer one. The energy-roofline model of Choi et al. \cite{choi2013energy,choi2014power} adds a third limit for power-capped processors: the total energy divided by the power available. When that limit binds, memory and compute compete for the same watts, so their times add instead of overlapping:

\[t\approx\frac{e_b B+e_f\Phi}{P}.\]
Here, \(e_b\) is the energy needed to move a unit of data, B is the number of bytes moved, \(e_f\) is the energy per unit of computation, \(\Phi\) is the total computation and P is the power available for use:

\[P=P_{\mathrm{cap}}-P_{\mathrm{static}}.\]
Here, \(P_{\mathrm{cap}}\) is the power cap of the GPU and \(P_{\mathrm{static}}\) is the minimum power consumed while the GPU is on. Over a full pass at 1,280 requests on H100 NVL, board power stays within 2\% of the 400 W cap in 96\% of 10 Hz samples, and the SM clock averages 1.32 GHz against a 1.785 GHz maximum.

\item Critical path: parallel branches finish with the longest, so shortening the longer branch provides the most direct reduction in completion time.

\item Filling the GPU: tiles must match the work; too few tiles leave SMs idle, which is why megakernels win below 384 tokens and 448 rows and larger encoder tiles lost above.

\end{enumerate}

\section{Evaluation}\label{sec:evaluation}

We evaluate end-to-end transcription on the English Bias corpus, comprising 4,820 recordings and 15.36 hours of audio, using single H100 SXM and H100 NVL GPUs with 700 W and 400 W power limits, respectively. Table~\ref{tab:h100} reports mean RTFx across three measured passes following a full-corpus warmup.

The vLLM baseline is NVIDIA's production configuration of vLLM 0.29.0 (vllm/vllm-openai:v0.29.0, build 98dff2a8) with the NeMo speech plugin and CUDA 13.0.2. It serves the ModelOpt FP8 checkpoint (FP8 linear layers, experts and KV cache) with 3-token MTP speculation. It uses FlashInfer attention~\cite{flashinfer}, CUTLASS dense GEMMs, FlashInfer-CUTLASS MoE and a BF16 Mamba state with 16-step ReplaySSM decoding. It also uses torch.compile with Inductor graph partitioning (decode sizes 512 to 768 precompiled), CUDA graphs up to 768 sequences, and the product's encoder graphs, packed Sortformer and head graphs. Limits are 768 concurrent sequences, 19,456 batched tokens per step, a 20 GiB KV-cache budget, 0.88 GPU memory utilization and four API servers. B200 uses the same limits with its SM100 FP8 profile. Both systems serve the same FP8 checkpoint on one GPU behind the same HTTP client. The client prepares all 4,820 requests before starting the clock, sends them in descending duration order with a 4 s ramp, and computes RTFx over the whole pass, including ramp and drain.

Our accuracy-preserving serving achieves 1.38 to 4.39\(\times\) higher throughput on SXM and 1.40 to 2.39\(\times\) on NVL, reaching 4,345\(\times\) and 3,381\(\times\) RTFx, or 1.39\(\times\) and 1.40\(\times\) vLLM's peak. At 128 concurrent requests, p95 end-to-end completion time decreases from 1.501 to 0.545 seconds on SXM and from 1.308 to 0.712 seconds on NVL. WER is 1.91 to 1.92\%, compared with 1.88 to 1.93\% for vLLM using the same checkpoint, workload, and text normalization; neither system is bit-exact, because batch composition changes FP8 GEMM tiling and reduction order.

\begin{table}[!htb]
\caption{End-to-end transcription throughput (RTFx) on H100.}\label{tab:h100}
\centering
\small
\setlength{\tabcolsep}{4pt}
\setlength{\arrayrulewidth}{0.5pt}
\renewcommand{\arraystretch}{1.08}
\setlength{\extrarowheight}{1pt}
\setlength{\tabcolsep}{3pt}
\begin{tabularx}{\linewidth}{|>{\raggedright\arraybackslash}p{.23\linewidth}|*{4}{>{\raggedright\arraybackslash}X|}}
\hline
\rowcolor{tableheader}
\textbf{RTFx} & \multicolumn{2}{c|}{\cellcolor{tableheader}\textbf{H100 SXM 700 W}} & \multicolumn{2}{c|}{\cellcolor{tableheader}\textbf{H100 NVL 400 W}} \\ \hline
\rowcolor{tableheader}
\textbf{Concurrent requests} & \textbf{vLLM} & \textbf{Optimized} & \textbf{vLLM} & \textbf{Optimized} \\ \hline
32 & 467 & 2,052 & 799 & 1,815 \\ \hline
64 & 788 & 2,990 & 1,025 & 2,451 \\ \hline
128 & 1,372 & 3,754 & 1,584 & 3,033 \\ \hline
256 & 2,241 & 4,181 & 2,054 & 3,316 \\ \hline
512 & 2,973 & 4,287 & 2,336 & 3,351 \\ \hline
768 & 3,127 & 4,324 & 2,410 & 3,369 \\ \hline
1,024 & 3,116 & 4,342 & 2,385 & 3,374 \\ \hline
1,280 & 3,020 & 4,345 & 2,402 & 3,381 \\ \hline
\end{tabularx}
\end{table}

vLLM is slower on H100 SXM than on H100 NVL below 256 requests because it is host-bound there: at 32 requests its mean GPU utilization is 45\% on the SXM host and 78\% on the NVL host, against 98\% for our engine on both. The hosts differ in CPU generation (Zen3 EPYC 7413 versus Zen4 EPYC 9124), CPU allocation and kernel; we did not isolate which of these slows vLLM's host path on SXM.

The regime-aware engine is 9\% faster than fused kernels alone at 8 requests, shrinking to 2\% at 1,280 as passes grow and 23\% faster than megakernels alone at 128 requests, growing to 32\% at 1,280. At 32 requests, where passes fall on both sides of the switch points, it beats both (5\% and 9\%). Even with fused kernels only, our engine is 4.2\(\times\) faster than vLLM at 32 requests and 1.41\(\times\) at 1,280; that gap comes from the native host path without Python, GPU feature extraction, fused epilogues and deeper speculation with CTC hints.

\begin{table}[!htb]
\caption{End-to-end transcription throughput (RTFx) on B200.}\label{tab:b200}
\centering
\small
\setlength{\tabcolsep}{4pt}
\setlength{\arrayrulewidth}{0.5pt}
\renewcommand{\arraystretch}{1.08}
\setlength{\extrarowheight}{1pt}
\begin{tabularx}{\linewidth}{|>{\raggedright\arraybackslash}p{.34\linewidth}|*{3}{>{\raggedright\arraybackslash}X|}}
\hline
\multicolumn{4}{|c|}{\cellcolor{tableheader}\textbf{RTFx on B200}} \\ \hline
\rowcolor{tableheader}
\textbf{Concurrent requests} & \textbf{vLLM} & \textbf{Optimized} & \textbf{Speedup} \\ \hline
32 & 870 & 2,675 & 3.07\(\times\) \\ \hline
64 & 1,041 & 4,011 & 3.85\(\times\) \\ \hline
128 & 1,591 & 5,124 & 3.22\(\times\) \\ \hline
256 & 2,565 & 5,958 & 2.32\(\times\) \\ \hline
512 & 3,549 & 6,220 & 1.75\(\times\) \\ \hline
768 & 3,777 & 6,297 & 1.67\(\times\) \\ \hline
1,024 & 3,850 & 6,350 & 1.65\(\times\) \\ \hline
1,280 & 3,894 & 6,367 & 1.64\(\times\) \\ \hline
\end{tabularx}
\end{table}

B200 runs a separate SM100 build of the same engine, not a recompile. Blackwell has no wgmma, so every tensor-core path in the megakernels, the draft head and the fused GEMMs is reimplemented with UMMA (tcgen05.mma), which accumulates in tensor memory. The SM100 code allocates TMEM, waits on MMA-completion barriers and drains accumulators before each epilogue; register-level mma.sync paths such as attention carry over. Persistent grids are sized from the SM count queried at start-up (144, not H100's 132), so the concurrent ASR and diarization megakernels split the GPU 72/72 instead of 94/38. The scheduler, switch points (384 tokens, 448 rows), admission limits (255 resident requests; 6,144 prompt tokens and 8,192 rows per step), MTP depth and numerical recipes are unchanged. Re-measured on B200, the crossovers stay within 32 tokens and 32 rows of the H100 values, so we keep the same thresholds. One B200-specific change mattered at high load. The draft's folded expert path now covers 256 rows instead of 128. That removed a slowdown once a draft pass exceeded 128 requests and raised throughput by 26 to 36\% from 256 concurrent requests up (4,688\(\times\) to 6,367\(\times\) at 1,280).

On NVIDIA B200 (Table~\ref{tab:b200}) (144 SMs, 1,000 W limit), our engine is 3.1\(\times\) faster than vLLM at 32 concurrent requests and 3.9\(\times\) at 64. The gap narrows to 1.6 to 1.8\(\times\) from 512 requests, where our engine levels off above 6,200\(\times\) real time (6,367\(\times\) and 555 requests/s at 1,280) while vLLM is still climbing toward 3,894\(\times\). The p95 end-to-end completion time is 1.9 to 3.8\(\times\) lower at every concurrency (0.21 s against 0.63 s at 32), and peak GPU memory is 56.0 GiB against 73.3 GiB. Our optimized serving holds WER at 1.92 to 1.93\% against vLLM's 1.89 to 1.91\%, within 0.04 percentage points. Enabling all opt-in optimizations adds 13 to 16\% throughput, reaching 7,414\(\times\) at 1,280 requests, but raises WER to 3.44 to 3.48\%.

With a 400 ms limit on p95 end-to-end completion time, our engine serves 64 concurrent requests on H100 SXM (0.365 s; 0.545 s at 128) and 128 on B200 (0.368 s; 0.595 s at 256), while vLLM meets the limit only up to 8 on H100 SXM and 16 on B200: 8\(\times\) as many requests within the budget. Our optimized serving on H100 SXM is more performant than vLLM serving on B200, bridging the hardware improvement over a GPU generation with software serving optimizations only.

We additionally evaluate the complete LibriSpeech~\cite{librispeech} test-clean and test-other splits: 5,559 recordings totaling 10.75 hours, with a mean duration of 6.96 seconds compared with 11.47 seconds for the Bias corpus. Requests follow a frozen shuffled order, and results are averaged over two measured passes after a full-corpus warmup, retaining the same end-to-end timing boundaries. Across 32 to 1,280 concurrent requests, our regime-aware serving achieves 1.60\(\times\) to 3.52\(\times\) vLLM's throughput on H100 SXM 700 W and 1.82\(\times\) to 3.90\(\times\) on B200, reaching 3,739\(\times\) and 5,219\(\times\) real time, respectively. At 128 concurrent requests, p95 end-to-end completion time decreases from 1.587 to 0.454 seconds on H100 SXM and from 1.356 to 0.328 seconds on B200. Combined WER is 2.437 to 2.457\%, compared with 2.449 to 2.485\% for vLLM. These results extend the throughput and latency benefits to a public corpus with shorter recordings and shuffled arrivals, supporting the applicability of the serving design beyond the original industrial workload.

\section{Related work}\label{sec:related}

Persistent kernels keep a computation resident across steps \cite{persistent-threads,persistent-rnns}, and recent LLM megakernels schedule tiles across operators in a single launch \cite{mpk2026,no-bubbles}, whereas path tracing found a monolithic megakernel harmful \cite{megakernels-harmful}. Following our earlier analysis \cite{chatterjee2026megakernel}, we run a megakernel only where a pass is small enough for it to win, and fused kernels above a measured crossover. Speculative decoding \cite{leviathan2023,chen2023speculative,eagle} drafts with a smaller model, extra decoding heads \cite{medusa2024} or multi-token prediction modules \cite{mtp2024,deepseekv3}; in hybrid models, verification must also restore the state-space states \cite{mamba-llama}.

Mixing prompt and decode rows in one step follows chunked prefill in Sarathi-Serve \cite{sarathi2024}, which vLLM also uses; we apply it to a hybrid model whose verification rows also need Mamba state replay. Megakernel serving engines \cite{cohere-megakernel,no-bubbles} mostly target decode; we target a prefill-dominated workload and switch per pass. RaMP \cite{ramp} dispatches MoE kernels from the live routing histogram, whereas we dispatch from pass size alone.

State-space and hybrid Mamba-Transformer models \cite{mamba,mamba2,nemotron-h} and speech LLMs such as Nemotron 3.5 Transcribe \cite{nemotron-transcribe} are usually served with stacks tuned for chat, while ASR serving has centered on batched inference for encoder-decoder models such as Whisper \cite{whisper}. We know of no published study of serving a prefill-dominated hybrid speech LLM with mixture-of-experts layers.

\section{Open problems and conclusion}\label{sec:conclusion}

Our regime-aware engine serves Nemotron 3.5 Transcribe at 2.74 times the audio per second of tuned vLLM at 128 concurrent requests on H100 SXM (1.38 to 4.39\(\times\) across 32 to 1,280), cuts p95 end-to-end completion time from 1.50 s to 0.55 s, and cuts GPU time per audio hour by 63\% at 128 requests and 28\% at peak; opt-in switches add 13 to 16\% on B200. We expect the lesson to carry to other hybrid models on power-capped GPUs: when the board draws its limit, speed comes from fewer bytes and fewer operations, not from better overlap.

This work raises three open problems:

\begin{enumerate}
\item The crossover points are determined offline. Different routing and admission limits can move the crossover points at runtime. What is the optimal online approach to determine crossover points?

\item Our approach of regime-aware megakernels and kernel fusion results in dynamic granularity at the operator level. Is this the optimal performance model for all megakernel-receding architectures?

\item Nemotron 3.5 Transcribe is a prefill-dominated megakernel-receding architecture. How do the optimization strategy and operator granularity change depending on different components becoming dominant?

\end{enumerate}

\FloatBarrier
\bibliographystyle{ACM-Reference-Format}
\bibliography{references}
\end{document}